\documentclass[preprint,preprintnumbers,amsmath,amssymb,superscriptaddress]{revtex4}

\usepackage{amsfonts}
\usepackage{subfig}
\usepackage{graphicx}
\usepackage{epsfig}
\usepackage{dcolumn}
\usepackage{bm}
\begin{document}
\title{Charged scalar production at the Compact Linear Collider for the $S_3 \otimes \mathbb{Z}_2$ model}

\author{G. De Conto}
\email{george.de.conto@gmail.com}
\affiliation{
Centro de Ci\^{e}ncias Naturais e Humanas \\ Universidade Federal do ABC,
09210-580 Santo Andr\'{e}-SP, Brasil}

\author{A. C. B. Machado}
\email{a.c.b.machado1@gmail.com}
\affiliation{
Centro de Ci\^{e}ncias Naturais e Humanas \\ Universidade Federal do ABC,
09210-580 Santo Andr\'{e}-SP, Brasil}

\author{J. Monta\~{n}o}
\email{jmontano@conacyt.mx}
\affiliation{CONACYT-Facultad de Ciencias F\'isico Matem\'aticas \\
Universidad Michoacana de San Nicol\'as de Hidalgo \\
58060, Morelia, Michoac\'an, M\'exico}

\author{P. Chimenti}
\email{pietro.chimenti@uel.br}
\affiliation{
	Departamento de F\'{i}sica, Centro de Ci\^{e}ncias Exatas\\
	Universidade Estadual de Londrina \\ Londrina - PR, 86051-990,
	Brazil}

\date{\today}

\begin{abstract}
	We present a model with $S_3 \otimes \mathbb{Z}_2$ model plus a sterile neutrino and its phenomenological expectations for the production of charged scalars at the Compact Linear Collider. At tree level, our model predicts a total cross section in between 0.1 and $10^{-5}$ pb for the $e^- e^+ \to H^+ H^-$ process, considering all possible mass values for the charged scalar in the CLIC experiment. We also show that this prediction holds regardless of the masses of the other exotic particles and their couplings.
\end{abstract}

\maketitle

\section{Introduction}
\label{sec:intro}

The Standard Model (SM) of fundamental interactions is still an incomplete theory, even though it was shown as an experimentally proven theory with a high predictive power, once we have observed dark matter (DM) from galaxy rotation curves, and there is no viable candidate for DM in the SM as it is today, the SM has an unknown missing piece. We also have the fact that neutrinos in the SM are non-massive, therefore, to have massive neutrinos, the only requirement is the addition of right handed (or sterile neutrino) fields to the SM matter content. However, because neutrinos are only charged under weak  isopsin and hyperchagre, some consequences arise because of the introduction of their right - handed components, for example, neutrinos can be a majorana particles. This means that neutrinos are the most elusive of the known particles. Their weak interactions make it very difficult to study their properties, but, this may be seen as a  good quality for the neutrinos, once they can hold the key to solve several mysteries in particle physics and cosmology. For instance, the simplest dark matter candidate, at least from the point of view of particle physics, is the neutrino.

The DM particle candidates may have very different masses (for reviews of DM candidates see e.g. \cite{9,10,11,12}): massive gravitons with mass $ \sim 10^{-19}$ eV \cite{13}, axions with mass $ \sim 10^{- 6}$ eV \cite{14}, sterile neutrinos having mass in the keV range \cite{Dodelson:1993je}, sypersymmetric (SUSY) particles (gravitinos \cite{16}, neutralinos \cite{17}, axinos \cite{18} with their masses ranging from eV to hundreds GeV, supersymmetric Q-balls \cite{19}, WIMPZILLAs with the mass $ \sim 10^{13}$ GeV \cite{20, 21}, and many others). Thus, the mass of DM particles becomes an important characteristic which may help to distinguish between various DM candidates and, more importantly, may help to differentiate among different models beyond the SM.

Sterile neutrinos as Dark Matter candidates, with the mass in the keV range, were originally suggested in~\cite{Dodelson:1993je}. Because these particles are neutral with respect to all SM charges, they have to be massive and, while unstable, they can have a lifetime longer than the age of the Universe (controlled by the active-sterile mixing parameter). If they exist, they were produced at high temperatures in the early Universe. The  sterile neutrinos are never  in thermal equilibrium, unlike other cosmic relic particles (e.g.\ photons, neutrinos or hypothetical WIMPs)  because of their feeble interaction, so their exact production mechanism is model-dependent. In Ref.\cite{Boyarsky:2018tvu} they  discuss the most important observational constraints on these particles as DM candidates.

Despite being a candidate for DM known for more than 15 years \cite{Dodelson:1993je}, recently the sterile neutrinos once again stood out as it was shown in Ref.\cite{49}. If three sterile right-handed neutrinos are added to the SM, it is possible to explain simultaneously the data on neutrino
oscillations (see e.g. \cite{50, 51, 52} for a review) and the Dark Matter in the Universe, without introducing any new physics above electro-weak scale.  Moreover, if the masses of two of these particles are between $\sim$ 100 MeV and the electro-weak scale, and are almost degenerate, it is also possible to generate the correct baryon asymmetry of the Universe (see e.g. \cite{49, 54, 55}). The third (lightest) sterile neutrino can have a mass in the keV-MeV range and be coupled to the rest of the matter weakly enough to provide a viable (cold or warm) DM candidate.

In this context, we have proposed an extension of the model with a $S_3\otimes \mathbb{Z}_2$ symmetry \cite{Dias:2012bh,Fortes:2017ndr} assuming that right-handed neutrinos can be sterile. The model consists of two scalar doublets and two scalar singlets plus the SM particles, with their potential having the most general lagrangian allowed by the chosen symmetry. We present a full analysis of the scalar potential of the model, showing the scalar mass eigenstates, and identify the SM Higgs boson among these by imposing that its Yukawa couplings are the same in our model and in the SM.

Models with a $S_3$ symmetry have been studied in a variety of previous works. Among these, some are: the model's scalar potential, including its mass eigenstates and self-couplings \cite{Kubo:2004ps,Barradas-Guevara:2014yoa,Das:2014fea}, the quark sector of these models \cite{Canales:2013cga}, and the neutrino masses and their mixing \cite{Canales:2012dr}. Despite some of these topics overlapping our work, we feel that what we present here is relevant. In our model two scalar singlets are added, which further increases the scalar sector and its complexity, leading to different mass eigenstates and Yukawa sectors. Therefore, our results are not the same as the ones shown in the works just cited.

Also, the extension of the scalar sector brings new charged scalars, which in turn provide new sources of CP violation, given that they come up in vertices with the form $\bar{f}_L f_R H^\pm$, where $f_{L,R}$ denotes left and right-handed fermions. While in the SM we have vertices with the form $\bar{f}_L f_L W^\pm$ that give rise to the CKM matrix, the $\bar{f}_L f_R H^\pm$ vertices mixes left and right-handed fermions, bringing in new combinations of matrices that diagonalize the fermion mass eigenstates. These new combinations of unitary matrices generate new sources of CP violation, which are necessary to solve other problems faced by the SM such as the observed matter--antimatter asymmetry of the Universe~\cite{Riotto:1999, Morrisey:2012, Dine:2004} and mass differences in mesonic systems \cite{Machado:2013jca,branco_Buchalla:2008jp}.

In order to explore the properties and phenomenological consequences of both type of particles from our model, charged scalars and sterile neutrinos, we are going to study the tree level process $e^- e^+ \to H^+ H^-$, where the sterile neutrinos participate as virtual particles and the charged scalars as final state particles. Because we are studying electron-positron collisions, we considered the Compact Linear Collider (CLIC), a multi TeV linear electron-positron collider under development \cite{Roloff:2018dqu,Robson:2018enq}, to calculate the cross section for the $e^- e^+ \rightarrow H^+ H^-$ process.

CLIC is a TeV-scale high-luminosity linear $e^+e^-$ collider under development by international collaborations hosted by CERN. For an optimal exploitation of its physics potential it is foreseen to be built and operated in stages at center-of-mass energies of 380 GeV, 1.5 TeV and 3 TeV, for a site length ranging between 11 km and 50 km. The high collision energy combined with the large luminosity and clean environment of the $e^+e^-$ collisions will enable to measure the properties of SM particles, such as the Higgs boson and the top quark, with unparalleled precision. CLIC might also discover indirect effects of very heavy new physics by probing the parameters of the Standard Model Effective Field Theory with an unprecedented level of precision. This includes new particles detected in non-standard signatures, such as electrically charged scalars. The construction of the first CLIC energy stage could start as early as 2026 and the first beams would be available by 2035, marking the beginning of a physics programme spanning $25-30$ years and providing excellent sensitivity to Beyond Standard Model physics, through direct searches and via a broad set of precision measurements of SM processes, particularly in the Higgs and top-quark sectors.

The outline of the paper is as follows: first we present the model and its Lagrangian, then we focus into its scalar sector calculating the mass eigenstates and indentify the SM Higgs boson within those, and finally calculate the cross-section for the $e^- e^+ \rightarrow H^+ H^-$ process in our model, followed by our conclusions.

\section{The $S_3 \otimes \mathbb{Z}_2$ model plus a Sterile Neutrino}
\label{sec:model}

Here we will use the  $S_3$ discrete symmetry in order to obtain a model with 3 Higgs doublets, being  two of them inert.  The $S_3$ symmetry consists of all permutations among three objects. However, the representation of order 3 is reducible and is decomposed in two irreducible representations: $\textbf{3}=\textbf{1}\oplus \textbf{2}$. Here we will write only the multiplications involving two doublets and two singlets (which will be used here for obtaining the Yukawa interactions) and the scalar potential that is invariant under the full symmetry, $SU(2)_L\otimes U(1)_Y\otimes S_3\otimes \mathbb{Z}_2$. Let $[x_1,x_2]$ and $[y_1,y_2]$ be two doublets of $S_3$, the multiplication $\textbf{2}\otimes\textbf{2}$ is given by
\begin{equation}
\left[\begin{array}{c} x_1 \\ x_2\end{array}\right]_2 \otimes \left[\begin{array}{c} y_1 \\y_2\end{array}\right]_2 = [x_1 y_1 + x_2 y_2 ]_1 + [x_1 y_2 - x_2 y_1 ]_{1^\prime} + \left[\begin{array}{c} x_1 y_2 + x_1 y_2 \\ x_1 y_1 - x_2 y_2\end{array}\right]_{2^\prime}=\textbf{1}\oplus\textbf{1}^\prime\oplus \textbf{2}^\prime,
\end{equation}
being $\textbf{1}$ and $\textbf{1}^\prime$ singlets and $\textbf{2}^\prime$ a doublet. Besides we have that $\textbf{1}\otimes\textbf{1}=\textbf{1}$ and $\textbf{1}^\prime\otimes\textbf{1}^\prime=\textbf{1}$. For more details about this and other discrete symmetries see Ref.~\cite{Ishimori:2010au}.  The $Z_N$ group, that is Abelian,  can be represented as discrete rotations, whose generators corresponds to a $2\pi/N$ rotation.

\begin{table}
		\centering
\begin{tabular}{|c|c|c|c|c|c|c|c|}
\hline
Symmetry & $L_{i}$ &  $l_{jR}$ & $N_s$ & $N_d$ & $S$ & $D$  & $\zeta_d$  \\ \hline
$S_{3}$  & 1 &  1  & 1 & 2 & 1 & 2  & 2  \\ \hline
$Z_3$ & $\omega$ & $\omega$ & $\omega$ & 1 & 1 & $\omega^2$ & $\omega$\\ \hline
\end{tabular}
\caption{Transformation properties of the fermion and scalar fields under $S_3$ and $\mathbb{Z}_2$ symmetries. Quarks and charged leptons are singlets of $S_3$ and even under $\mathbb{Z}_2$.}
\label{table1}
\end{table}

The Yukawa interactions are given by
\begin{eqnarray}
	\label{yukawa}
	-\mathcal{L}^{leptons}_{Yukawa}&=& G^l_{ij}\bar{L}_{i}l_{jR}S+G^\nu_{ij} L_{i} \nu_{j L} S +  G^N_{is}L_{i}\epsilon N_{sR} S    \\ \nonumber&+& \frac{y_d}{2}[[\overline{N^c_{d R}} N_{d R}]_2 \zeta_d]_1 + \frac{y_{sd}}{2}[\overline{N^c_{s R}} [N_{d R} \zeta_d]_2]_1 + \frac{F^\nu_{is}}{\Lambda} \bar{L^c}_{i} N_{s} [D_d \zeta_d^*]_1 + H.c.,
\end{eqnarray}
where $i,j=e,\mu,\tau$ and $d=1,2$. In Eq. \ref{yukawa} $L$ is the SM lepton doublet, $l$ is the SM charged lepton, $\nu$ is the SM neutrino, $D_{1,2}$ and $S$ are scalar SU(2) doublets, $N_s$ and $N_{1,2}$ are right-handed neutrinos. $G^l$, $G^\nu$, $G^N$, $y_d$, $y_{sd}$ and $F^\nu$ are matrices for the Yukawa couplings and $\Lambda$ is an energy scale above the electroweak scale for the dimension 5 operator. For more details about the model see \cite{Fortes:2017ndr}, the difference in our work is that we changed the interactions of $N_s$ to make it lighter than the other right-handed neutrinos.

The scalar  SU(2) doublets, $D_{1,2}$ and $S$, are introduced to generate Majorana masses for the left-handed neutrinos, according to the scotogenic model. These doublets may also be candidates for dark matter, a possibility that was studied in \cite{Fortes:2017ndr}. In order to have a lighter neutrino than the scale of the heavier Majorana neutrinos we use the $S_3$ symmetry, that separates the right-hand neutrinos into a doublet and singlet (this lighter neutrino may fit the evidences of lighter sterile neutrinos). However, for the symmetry to be conserved it is necessary to introduce two scalar singlets ($\zeta_{1,2}$), so that the heavier right-handed neutrinos have Majorana masses on the TeV scale.

\section{Scalar sector}
\label{sec:scalars}

The scalar sector of the model is presented as follows:
\begin{equation}
S=\left (\begin{array}{c}
S^+ \\
\frac{1}{\sqrt2}(v_{SM}+\textrm{Re}S^0+i\textrm{Im}S^0)
\end{array}\right),\quad D_{1,2}=\left( \begin{array}{c}
D^+_{1,2} \\
\frac{1}{\sqrt2}(\eta_{1,2}+i\chi_{1,2}).
\end{array}\right),
\label{notation}
\end{equation}
plus two real the singlets $\zeta_i=\frac{v_i+\xi_i}{\sqrt{2}}$, $i=1,2$.

The scalar potential invariant under the gauge and $S_3\otimes \mathbb{Z}_2$ symmetries is
\begin{eqnarray}
	V_{S_3}  &=& \mu^2_sS^\dagger S+\mu^2_d [D^\dagger\otimes  D]_1+ \mu^2_\zeta [\zeta_d\otimes  \zeta_d]_1 +\mu^2_{12}\zeta_1\zeta_2+a_1
	([D^\dagger\otimes  D]_1)^2
	+  a_2 [[D^\dagger\otimes D]_{1^\prime}[D^\dagger\otimes
	D]_{1^\prime}]\nonumber\\
	&&+a_3[(D^\dagger \otimes D)_{2^\prime}(D^\dagger\otimes D)_{2^\prime}]_1
	+a_4(S^\dagger S)^2+
	a_5[D^\dagger\otimes D]_1 S^\dagger  S + a_6 [[S ^\dagger D]_{2^\prime} [S^\dagger   D]_{2^\prime}]_1
	\nonumber \\
	&&+ H.c.]+
	a_7 S^\dagger [ D \otimes D^\dagger]_1 S+
	b_1 S^\dagger S [ \zeta_d \otimes \zeta_d]_1 + b_2  [D^\dagger\otimes  D]_1 [ \zeta_d \otimes \zeta_d]_1
	+ b_3 [[D^\dagger\otimes  D]_{2^\prime} [ \zeta_d \otimes \zeta_d]_{2^\prime}]_{1} \nonumber\\&&+b_4 [[[D^\dagger\otimes D]_{1^\prime}[\zeta_d\otimes\zeta_d]_{1^\prime}]_1
	+  c_1([\zeta_d\otimes  \zeta_d]_1)^2+
	c_2 [[\zeta_d\otimes \zeta_d]_{2^\prime}[\zeta_d\otimes
	\zeta_d]_{2^\prime}]_1,
	\label{potential1}
\end{eqnarray}
with $\mu^2_d>0$ since $\langle D^0_{1,2}\rangle=0$ is guaranteed by the $S_3$ symmetry. The parameter $a_6$ has been chosen real without loss of generality.

We can write Eq.~(\ref{potential1}) explicitly as
\begin{equation}
V(S,D,\zeta_d)=V^{(2)}+V^{(4a)}+V^{(4b)}+V^{(4c)},
\label{potential2}
\end{equation}
where
\begin{eqnarray*}
	V^{(2)}&=&\mu^2_{SM}S^\dagger S+\mu^2_d (D^\dagger_1D_1+D^\dagger_2 D_2)+\mu^2_\zeta(\zeta^2_1+\zeta^2_2)+\mu^2_{12}\zeta_1\zeta_2,
	\nonumber\\
	V^{(4a)} &=&  a_1
	(D^\dagger_1D_1+D^\dagger_2 D_2)^2
	+  a_2 (D^\dagger_1D_2-D^\dagger_2D_1)^2
	\nonumber\\
	&&+ a_3[(D^\dagger_1D_2+D^\dagger_2D_1)^2+(D^\dagger_1D_1-D^\dagger_2D_2)^2]
	+a_4(S^\dagger S)^2+
	a_5  (D^\dagger_1D_1+D^\dagger_2 D_2)S^\dagger  S
	\nonumber\\
	&&+ a_6[(S^\dagger D_1)^2 +(S^\dagger D_2)^2 +H.c.]
	+
	a_7 [(S^\dagger D_1)(D^\dagger_1S)+(S^\dagger D_2)(D^\dagger_2S)],
	\nonumber\\
	V^{(4b)}&=& b_1S^\dagger S(\zeta^2_1+\zeta^2_2)+b_2(D^\dagger_1D_1+D^\dagger_2D_2)(\zeta^2_1+\zeta^2_2)+
	b_3[(D^\dagger_1D_2+D^\dagger_2D_1)(\zeta_1\zeta_2+\zeta_1\zeta_2)
	\nonumber \\
	&&+ (D^\dagger_1D_1-D^\dagger_2D_2)(\zeta^2_1-\zeta^2_2)+H.c.] +  b_4[(D^\dagger_1D_2-D^\dagger_2D_1)(\zeta_1\zeta_2-\zeta_1\zeta_2)]
	\nonumber\\
	V^{(4c)}&=&c_1(\zeta^2_1+\zeta^2_2)^2
	+ c_2[(\zeta_1\zeta_2+\zeta_2\zeta_1)^2+(\zeta^2_1-
	\zeta^2_2)^2],
	\label{potential3}
\end{eqnarray*}
where we have used $[\zeta_d\,\zeta_d]_{2^\prime}=(\zeta_1\zeta_2+\zeta_2\zeta_1,\zeta_1\zeta_1-\zeta_2\zeta_2)$. Here, we will consider all the couplings to be real parameters i.e., there is no $C\!P$ violation in the scalar sector.
The $S_3$ symmetry forbids linear terms with the doublets $D_1,D_2$ in the scalar potential and also some of the Yukawa interactions with charged leptons. This ensures the inert character of the these doublets after the $S_3$ symmetry is introduced. Notice that, although the term $\mu^2_{12}$ breaks softly the $S_3$ symmetry, it happens in the sector of the singlets $\zeta_{1,2}$ and does not spoil the inert character of the doublets.

From Eq.~(\ref{potential1}), we obtain the following stability conditions for the potential (i.e., setting its derivatives to zero):
\begin{eqnarray}
	\frac{1}{2} v_{SM} \left(2 a_4 v_{SM}^2+b_1 \left(v_1^2+v_2^2\right)+2 \mu_{SM}^2\right)&=&0,\nonumber \\
	\frac{1}{2} \left(b_1 v_1 v_{SM}^2+2 v_1 (c_1+c_2) \left(v_1^2+v_2^2\right)+\mu_{12}^2 v_2\right)+\mu_\zeta^2 v_1&=&0, \label{ce1} \\
	\frac{1}{2} b_1 v_2 v_{SM}^2+v_2 (c_1+c_2) \left(v_1^2+v_2^2\right)+\frac{\mu_{12}^2 v_1}{2}+\mu_\zeta^2 v_2&=&0, \nonumber
	\end{eqnarray}
From Eq. \ref{ce1} we find three sets of solutions
\begin{equation}
v_1=v_2=0, \quad \mu_{SM}=-a_4 v_{SM}^2
\label{eq:solderivadas1}
\end{equation}
\begin{equation}
v_2= -v_1,\quad \mu_\zeta^2= \frac{1}{2} \left(-b_1 v_{SM}^2-4 v_1^2 (c_1+c_2)+\mu_{12}^2\right), \quad \mu_{SM}=-a_4 v_{SM}^2-b_1 v_1^2
\label{eq:solderivadas2}
\end{equation}
\begin{equation}
v_2= v_1,\quad \mu_\zeta^2= \frac{1}{2} \left(-b_1 v_{SM}^2-4 v_1^2 (c_1+c_2)-\mu_{12}^2\right), \quad \mu_{SM}=-a_4 v_{SM}^2-b_1 v_1^2
\label{eq:solderivadas3}
\end{equation}
These solutions will be used in the analyses presented throughout this work.

\section{Mass matrices and eigenstates}\label{sec:massmatrices}

The potential in Eq. \ref{potential1} gives us four mass matrices: one for the charged scalars, one for the CP-odd neutral scalars and two for the CP-even neutral scalars. In the sections below we show these matrices and their corresponding eigenvalues and eigenvectors. Also, when we have $\pm $ or $\mp$, the upper sign corresponds to the vacuum stability criteria from Eq. \ref{eq:solderivadas2} and the lower sign to Eq. \ref{eq:solderivadas3}.

\subsection{Charged scalars}

From Eq. \ref{potential2}, in the basis $(S^+, D_1^+, D_2^+) M_C (S^-,D_1^-,D_2^-)^T$, we find the mass matrix $M_C$ for the charged scalars to be
\begin{equation}
M_C=
\left(
\begin{array}{ccc}
0 & 0 & 0 \\
0 & b_2 v_1^2+\frac{a_5 v_{SM}^2}{2}+\mu_d^2 & \mp \, 2 b_3 v_1^2 \\
0 & \mp \, 2 b_3 v_1^2 & b_2 v_1^2+\frac{a_5 v_{SM}^2}{2}+\mu_d^2 \\
\end{array}
\right).
\label{eq:matmassacarregados}
\end{equation}
The above matrix can be diagonalized as $R_C^T M_C R_C$, where $R_C$ is the orthogonal rotation matrix. For the matrix in Eq. \ref{eq:matmassacarregados} we find that the symmetry and mass eigenstates are related as
\begin{equation}
\left(
\begin{array}{c}
S^+ \\ D_1^+ \\ D_2^+
\end{array}
\right)
= R_C
\left(
\begin{array}{c}
G^+ \\ H_1^+ \\ H_2^+
\end{array}
\right)
=
\left(
\begin{array}{ccc}
1 & 0 & 0 \\
0 & \pm \, \frac{1}{\sqrt{2}} & \mp \, \frac{1}{\sqrt{2}} \\
0 & \frac{1}{\sqrt{2}} & \frac{1}{\sqrt{2}} \\
\end{array}
\right)
\left(
\begin{array}{c}
G^+ \\ H_1^+ \\ H_2^+
\end{array}
\right).
\end{equation}
And their masses are:
\begin{equation}
m_{G^+}^2=0, \quad m_{H_1^+}^2=\frac{a_5 v_{SM}^2}{2}+v_1^2 (b_2-2 b_3)+\mu_d^2, \quad m_{H_2^+}^2=\frac{a_5 v_{SM}^2}{2}+v_1^2 (b_2+2 b_3)+\mu_d^2.
\end{equation}

\subsection{CP-odd scalars}

From Eq. \ref{potential2}, in the basis $(Im[S^0],\chi_1,\chi_2) M_O (Im[S^0],\chi_1,\chi_2)^T$, we find the mass matrix $M_O$ for the CP-odd scalars to be
\begin{equation}
M_O=
\left(
\begin{array}{ccc}
0 & 0 & 0 \\
0 &  b_2   v_1 ^2+\frac{1}{2} ( a_5 -2  a_6 + a_7 )  v_{SM} ^2+ \mu_d^2  & \mp 2  b_3   v_1 ^2 \\
0 & \mp 2  b_3   v_1 ^2 &  b_2   v_1 ^2+\frac{1}{2} ( a_5 -2  a_6 + a_7 )  v_{SM} ^2+ \mu_d^2  \\
\end{array}
\right).
\end{equation}
The diagonalization for this mass matrix is given by
\begin{equation}
\left(
\begin{array}{c}
Im[S^0] \\ \chi_1 \\ \chi_2
\end{array}
\right)
=
R_O
\left(
\begin{array}{c}
G_A^0  \\ A_1^0 \\ A_2^0
\end{array}
\right)
=
\left(
\begin{array}{ccc}
1 & 0 & 0 \\
0 & \pm \, \frac{1}{\sqrt{2}} & \mp \, \frac{1}{\sqrt{2}} \\
0 & \frac{1}{\sqrt{2}} & \frac{1}{\sqrt{2}} \\
\end{array}
\right)
\left(
\begin{array}{c}
G_A^0  \\ A_1^0 \\ A_2^0
\end{array}
\right).
\end{equation}
With masses:
\begin{align}
m_{G_A^0}^2=0, \quad m_{A_1^0}^2=\frac{1}{2}  v_{SM} ^2 ( a_5 -2  a_6 + a_7 )+ v_1 ^2 ( b_2 -2  b_3 )+ \mu_d^2 ,\\ m_{A_2^0}^2=\frac{1}{2}  v_{SM} ^2 ( a_5 -2  a_6 + a_7 )+ v_1 ^2 ( b_2 +2  b_3 )+ \mu_d^2 . \nonumber
\end{align}

\subsection{CP-even scalars, $2\times 2$ matrix}

Eq. \ref{potential2} gives us for the CP-even sector two matrices, one $2\times 2$ and one $3 \times 3$. The $2\times 2$ matrix, in the basis $(\eta_1,\eta_2) M_{E1} (\eta_1,\eta_2)^T$, is
\begin{equation}
M_{E1}=
\left(
\begin{array}{cc}
b_2   v_1 ^2+\frac{1}{2} ( a_5 +2  a_6 + a_7 )  v_{SM} ^2+ \mu_d^2  & \mp 2  b_3   v_1 ^2 \\
\mp 2  b_3   v_1 ^2 &  b_2   v_1 ^2+\frac{1}{2} ( a_5 +2  a_6 + a_7 )  v_{SM} ^2+ \mu_d^2  \\
\end{array}
\right).
\end{equation}
The symmetry and mass eigenstates are related as
\begin{equation}
\left(
\begin{array}{c}
\eta_1 \\ \eta_2
\end{array}
\right)
=
R_{E1}
\left(
\begin{array}{c}
h^0_1 \\ h^0_2
\end{array}
\right)
=
\left(
\begin{array}{cc}
\pm \, \frac{1}{\sqrt{2}} & \mp \, \frac{1}{\sqrt{2}} \\
\frac{1}{\sqrt{2}} & \frac{1}{\sqrt{2}} \\
\end{array}
\right)
\left(
\begin{array}{c}
h_1^0 \\ h_2^0
\end{array}
\right).
\end{equation}
The masses are:
\begin{align}
m_{h_1^0}^2=\frac{1}{2}  v_{SM} ^2 ( a_5 +2  a_6 + a_7 )+ v_1 ^2 ( b_2 -2  b_3 )+ \mu_d^2 , \\ m_{h_2^0}^2=\frac{1}{2}  v_{SM} ^2 ( a_5 +2  a_6 + a_7 )+ v_1 ^2 ( b_2 +2  b_3 )+ \mu_d^2 . \nonumber
\end{align}

\section{The CP-even $3 \times 3$ matrix and the Higgs boson}\label{sec:CPpar3x3}

Again from Eq. \ref{potential2}, we obtain the $3 \times 3$ matrix for the CP-even scalars. Considering the basis $(Re[S^0], \xi_1,\xi_2) M_{E2} (Re[S^0], \xi_1, \xi_2)$ we find
\begin{equation}
M_{E2}=
\left(
\begin{array}{ccc}
2  a_4   v_{SM} ^2 & b_1   v_1   v_{SM}  & \mp b_1   v_1   v_{SM}  \\
b_1   v_1   v_{SM}  & \frac{1}{2} \left(4 ( c_1 + c_2 )  v_1 ^2 \pm \mu_{12}^2 \right) & \frac{1}{2} \left( \mu_{12}^2 \mp 4 ( c_1 + c_2 )  v_1 ^2\right) \\
\mp b_1   v_1   v_{SM}  & \frac{1}{2} \left( \mu_{12}^2 \mp 4 ( c_1 + c_2 )  v_1 ^2\right) & \frac{1}{2} \left(4 ( c_1 + c_2 )  v_1 ^2 \pm \mu_{12}^2 \right) \\
\end{array}
\right),
\label{eq:MatMassaCPpar3x3}
\end{equation}
where, when we have $\pm$ or $\mp$, the upper sign corresponds to the vacuum stability criteria from Eq. \ref{eq:solderivadas2} and the lower sign to Eq. \ref{eq:solderivadas3}. To find the mass eigenstates of this sector, we will follow Ref. \cite{Higgsm331}, where we impose that the Yukawa couplings of the Higgs boson are the same in the SM and in the $S_3\otimes \mathbb{Z}_2$ model.

In the $S_3\otimes \mathbb{Z}_2$ model, the Yukawa sector for the leptons is given by
\begin{eqnarray}
\label{eq:Yukawa}
-\mathcal{L}^{leptons}_{Yukawa}&=& G^l_{ij}\bar{L}_{i}l_{jR}S+G^\nu_{ij} L_{i} \nu_{j L} S +  G^N_{is}L_{i}\epsilon N_{sR} S    \\ \nonumber&+& \frac{y_d}{2}[[\overline{N^c_{d R}} N_{d R}]_2 \zeta_d]_1 + \frac{y_{sd}}{2}[\overline{N^c_{s R}} [N_{d R} \zeta_d]_2]_1 + \frac{F^\nu_{is}}{\Lambda} \bar{L^c}_{i} N_{s} [D_d \zeta_d^*]_1 + H.c.,
\end{eqnarray}
where$i, j = e, \mu, \tau$, $d = 1,2$ (we omit summation symbols), $L_i(l_{iR})$ denote the usual left-handed lepton doublets (right-handed charged lepton singlets), $G's$ are the Yukawa couplings, and $N_{s,d}$ are the right-handed neutrinos.

From Eq. \ref{eq:Yukawa}, $Re[S^0]$ is the only one that gives mass to the known fermions, therefore we will identify it as the SM Higgs. The matrix in Eq. \ref{eq:MatMassaCPpar3x3} can be diagonalized by an orthogonal $3\times3$ matrix, such that $R^T_{E2} M_{E2} R_{E2}=diag(m^2_H, m^2_{H^0_1}, m^2_{H^0_2})$, where $H$ is the SM Higgs boson and $H^0_{1,2}$ are the other CP-even scalars from the $3 \times 3$ CP-even mass matrix. This implies
\begin{align}
\left(\begin{array}{c}
H \\ H_1^0 \\ H_2^0
\end{array}\right)
&=R^T_{E2}
\left(
\begin{array}{c}
Re[S^0] \\ \xi_1 \\ \xi_2
\end{array}
\right)
\\&=
\left(
\begin{array}{ccc}
cos \theta_2 & -cos \theta_3 sin \theta_2 & sin \theta_2 sin \theta_3 \\
cos \theta_1 sin \theta_2 & cos \theta_1 cos \theta_2 cos \theta_3-sin \theta_1 sin \theta_3 & -cos \theta_3 sin \theta_1-cos \theta_1 cos \theta_2 sin \theta_3 \\
sin \theta_1 sin \theta_2 & cos \theta_2 cos \theta_3 sin \theta_1+cos \theta_1 sin \theta_3 & cos \theta_1 cos \theta_3-cos \theta_2 sin \theta_1 sin \theta_3 \\
\end{array}
\right)
\left(
\begin{array}{c}
Re[S^0] \\ \xi_1 \\ \xi_2
\end{array}
\right). \nonumber
\end{align}
Since $Re[S^0]$ is the scalar we identify as the SM Higgs, we need that $(R^T_{E2})_{11}=1$ and all the other elements from the first row to be zero. To do so, we need $\theta_2=0$, which gives us $cos \theta_2=1$ and $sin \theta_2=0$, leaving $R_{E2}$ as
\begin{equation}
R_{E2}=
\left(
\begin{array}{ccc}
1 & 0 & 0 \\
0 & cos \theta_1 cos \theta_3-sin \theta_1 sin \theta_3 & cos \theta_3 sin \theta_1+cos \theta_1 sin \theta_3 \\
0 & -cos \theta_3 sin \theta_1-cos \theta_1 sin \theta_3 & cos \theta_1 cos \theta_3-sin \theta_1 sin \theta_3 \\
\end{array}
\right)
=
\left(
\begin{array}{ccc}
1 & 0 & 0 \\
0 & cos \theta & sin \theta \\
0 & -sin \theta & cos \theta \\
\end{array}
\right),
\label{eq:RotacaoCPpar3x3}
\end{equation}
where $\theta=\theta_1+\theta_3$. The matrix $R_{E2}$ from Eq. \ref{eq:RotacaoCPpar3x3} does not automatically diagonalize $M_{E2}$, i.e., it does not lead to $R^T_{E2} M_{E2} R_{E2}=diag(m^2_H, m^2_{H^0_1}, m^2_{H^0_2})$. To have that we need to impose conditions on the parameters that make up $M_{E2}$ and $R_{E2}$, so that we fulfill the equation $R^T_{E2} M_{E2} R_{E2}=diag(m^2_H, m^2_{H^0_1}, m^2_{H^0_2})$. The possible solutions depend on which solution for the vacuum stability conditions (Eqs. \ref{eq:solderivadas1}-\ref{eq:solderivadas3}) we choose.
\begin{enumerate}
	\item $v_1=v_2=0$:
	\subitem This solution trivializes our mass matrix $M_{E2}$, making it diagonal from the very beginning, leaving one scalar with mass $m^2_{Re[S^0]}=a_4 v_{SM}^2$ and the other two scalars with mass $m^2_{\xi_1,\xi_2}=-\mu_{12}^2/4$. Therefore, there is no need for the diagonalization method presented in this section.
	
	\item $ v_2= -v_1$ and $\mu_\zeta^2= \frac{1}{2} \left(-b_1 v_{SM}^2-4 v_1^2 (c_1+c_2)+\mu_{12}^2\right)$:
	\subitem In this case we find 10 possible solutions for the parameters $a_4$, $b_1$, $\mu_{12}^2$ and the sines and cosines of $\theta$. Amongst all solutions, we either have all masses equal, $m^2=4 v_1^2 (c_1+c_2)$; or two equal masses, $m^2_1=2 a_4 v_{SM}^2$, and a third different mass $m^2_2=4 v_1^2(c_1+c_2)$. In both cases, $b_1=0$. When all masses are equal, $a_4=\frac{2 v_1^2 (c_1+c_2)}{v_{SM}^2}$ and $\mu_{12}^2=4 v_1^2 (c_1+c_2)$; when we have different masses, $\mu_{12}^2=2 a_4 v_{SM}^2$. As for the sines and cosines, they can either be functions of each other or have values $\pm \, 1/\sqrt{2}$.
	
	\item $ v_2= v_1,\quad \mu_\zeta^2= \frac{1}{2} \left(-b_1 v_{SM}^2-4 v_1^2 (c_1+c_2)-\mu_{12}^2\right)$:
	\subitem Once again we have 10 possible solutions for the parameters $a_4$, $b_1$, $\mu_{12}^2$ and the sines and cosines of $\theta$. Amongst all solutions, we either have all masses equal, $m^2=4 v_1^2 (c_1+c_2)$; or two equal masses, $m^2_1=2 a_4 v_{SM}^2$, and a third different mass $m^2_2=4 v_1^2 (c_1+c_2)$. In both cases, $b_1=0$. When all masses are equal, $a_4=\frac{2 v_1^2 (c_1+c_2)}{v_{SM}^2}$ and $\mu_{12}^2=-4 v_1^2 (c_1+c_2)$; when we have different masses, $\mu_{12}^2=-2 a_4 v_{SM}^2$. As for the sines and cosines, they can either be functions of each other or have values $\pm \, 1/\sqrt{2}$.
\end{enumerate}
The complete set of solutions are shown in Appendix \ref{sec:SolucoesCPpar3x3}. 

\section{The $e^- e^+ \rightarrow H^+ H^-$ process at CLIC}
\label{sec:ilc}

\begin{figure}
\includegraphics[width=0.7\textwidth]{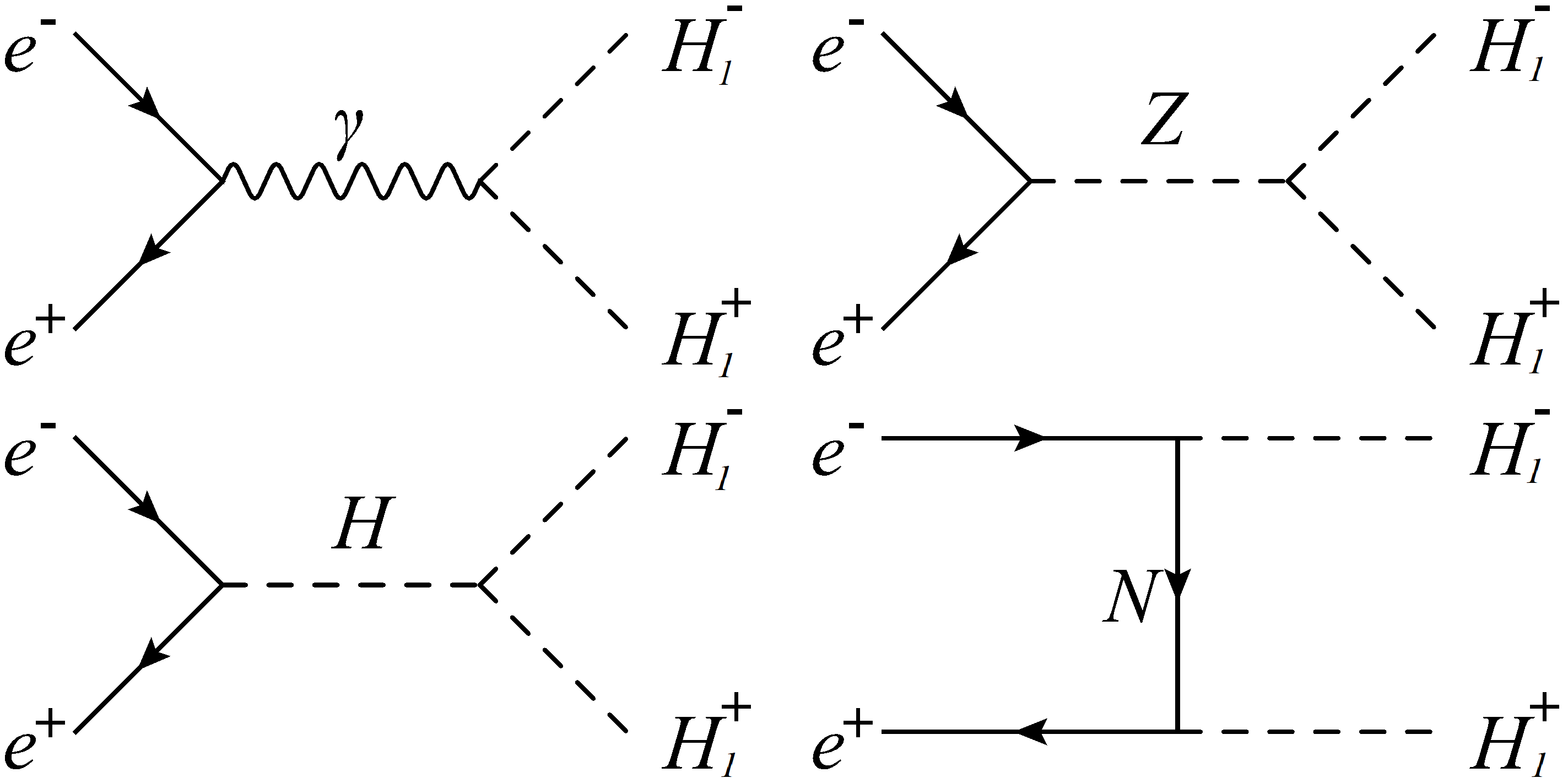}
\centering
\caption{Lowest order contributions to the $e^- e^+ \rightarrow H^+ H^-$ process.}
\label{fig:diagramas}
\end{figure}

The lowest order contributions to the $e^-e^+\rightarrow H^+ H^-$ process in our model are given by the diagrams in Fig. \ref{fig:diagramas}. For the scalar mass eigenstates we considered solution 10 from the appendix \ref{sec:SolucoesCPpar3x3v1=v2}. The model has two charged scalars, $H_1^\pm$ and $H_2^\pm$. In here we assumed that $H_1^\pm$ is the lightest one, and that is the one considered in our cross sections. Taking these assumptions into account, the relevant vertices for our calculations are:
\begin{equation}
H^+ H^- Z \rightarrow - \sqrt{\pi \alpha} \frac{c_W^2-s_W^2}{c_W s_W},
\end{equation}
\begin{equation}
H^+ H^- \gamma \rightarrow \sqrt{4 \pi \alpha},
\end{equation}
\begin{equation}
e^\pm N H^\pm \rightarrow \sqrt{2} F^\nu_{1S} \frac{v_1}{\Lambda}=\sqrt{2} X,
\label{eq:verticeeHPHM}
\end{equation}
\begin{equation}
H H^+ H^- \rightarrow a_5 v_{SM},
\end{equation}
\begin{equation}
e^+ e^- H \rightarrow \frac{m_e}{v_{SM}};
\end{equation}
where $\alpha$ is the fine structure constant, $v_{SM}$ is the SM vacuum expectation value for the Higgs field, $m_e$ is the electron mass, and $c_W$ and $s_W$ are the cosine and sine of the electroweak angle, respectively. $F^\nu_{1S}$ and $\Lambda$ are parameters of the Yukawa sector of our model (Eq. \ref{yukawa}), and $v_1$ and $a_5$ are parameters of the scalar sector of our model.

To calculate the cross sections we considered the three energy stages that CLIC will run at: 380 GeV, 1.5 TeV and 3 TeV (see Table \ref{tab:enegiasCLIC}). Throughout this section we will consider $X\equiv v_1F_{1s}^\nu/\Lambda=$ 0.1 or 0.01 (these values for $X$ allow perturbative calculations). Also, we will consider $a_5=1$ for the coupling constant in the $HH_1^+H_1^-$ vertex, the highest possible value that allows perturbative calculations.

\begin{table}[h]
	\begin{center}
	\begin{tabular}{|c|c|c|}
		\hline
		Stage & $\sqrt{s}$ (TeV) & $\mathcal{L}_{int}$ $(ab^{-1})$ \\
		\hline
		1 & 0.38 & 1.0 \\
		2 & 1.5 & 2.5 \\
		3 & 3 & 5.0 \\
		\hline
	\end{tabular}
\caption{Baseline CLIC energy stages and integrated luminosities for each stage \cite{Roloff:2018dqu}.}
\label{tab:enegiasCLIC}
	\end{center}
\end{table}

\begin{figure}[h]
\centering
\subfloat[]{\includegraphics[width=7.25cm]{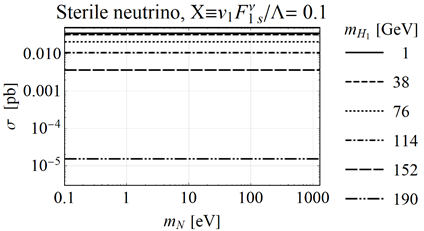}} \
\subfloat[]{\includegraphics[width=7.50cm]{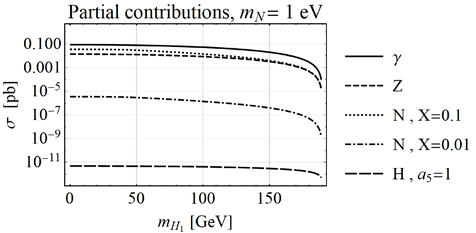}} \\
\subfloat[]{\includegraphics[width=7.25cm]{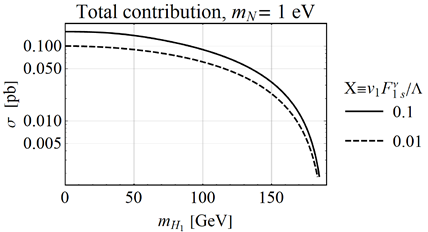}}
\caption{Predictions for the $e^+ e^-\to H^+_1H^-_1$ cross section with $\sqrt{s}=$ 380 GeV. (a): The sterile neutrino contribution for the cross-section as a function of its mass, considering several mass values for the charged scalar. (b): The partial contribution to the cross section as a function of the charged scalar mass. (c): The total cross section as a function of the charged scalar mass.}
\label{fig:SecoesChoque380GeV}
\end{figure}

In Fig. \ref{fig:SecoesChoque380GeV} are presented the $e^+ e^-\to H^+_1H^-_1$ cross section predictions for $\sqrt{s}=$ 380 GeV. The top left graph shows that the contribution of the sterile neutrino to the cross section is constant in the range $m_N=[1,1000]$ eV, therefore we set $m_N=1$ eV in the other plots, which is compatible with the current experimental data \cite{PDG}. In the top right graph the partial contributions show the hierarchy of the contributors: the photon is the main responsible, it is followed by the sterile neutrino with $m_N=1$ eV and $X=0.1$, the third place is due to the $Z$ gauge boson; then the signal of the  neutrino with $X=0.01$ is suppressed, and the most marginal participation comes from the Higgs boson. At the bottom we show the total cross section, taking into account all diagrams from Fig. \ref{fig:diagramas}. One can notice that for $0\,GeV\,<m_{H_1}<130\,GeV$ the total cross section stays in the interval $\sigma=$[1 pb, 0.005 pb].

\begin{figure}
\centering
\subfloat[]{\includegraphics[width=7.25cm]{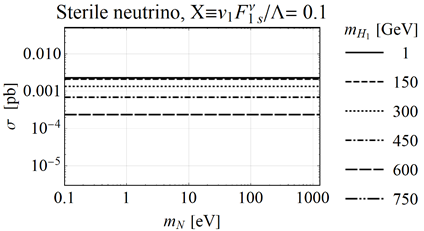}} \
\subfloat[]{\includegraphics[width=7.50cm]{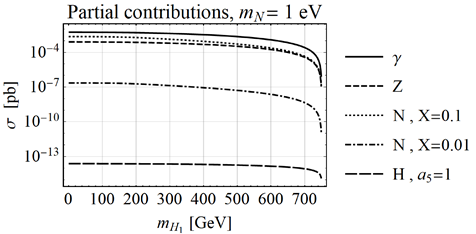}} \\
\subfloat[]{\includegraphics[width=7.25cm]{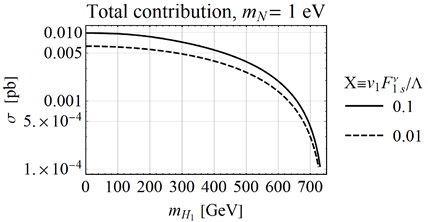}}
	\caption{Predictions for the $e^+ e^-\to H^+_1H^-_1$ cross section with $\sqrt{s}=$ 1.5 TeV. (a): The sterile neutrino contribution for the cross-section as a function of its mass, considering several mass values for the charged scalar. (b): The partial contribution to the cross section as a function of the charged scalar mass. (c): The total cross section as a function of the charged scalar mass.}
	\label{fig:SecoesChoque1500GeV}
\end{figure}

In Fig. \ref{fig:SecoesChoque1500GeV} are presented the $e^+ e^-\to H^+_1H^-_1$ cross section predictions for $\sqrt{s}=$ 1.5 TeV. The top left graph shows that the contribution of the sterile neutrino to the cross section is once again constant in the range $m_N=[1,1000]$ eV, therefore we set $m_N=1$ eV in the other plots. In the top right graph the partial contributions follow the same hierarchy from Fig. \ref{fig:SecoesChoque380GeV}, this time however, the cross sections are about two orders of magnitude smaller than before. At the bottom we show the total cross section, taking into account all diagrams from Fig. \ref{fig:diagramas}. One can notice that for $0$ GeV$<m_{H_1}<700$ GeV the total cross section stays in the interval $\sigma=$[0.01 pb, 0.0001 pb], once again about two orders of magnitude smaller than in the $\sqrt{s}=$ 380 GeV case.

\begin{figure}[h]
\centering
\subfloat[]{\includegraphics[width=7.25cm]{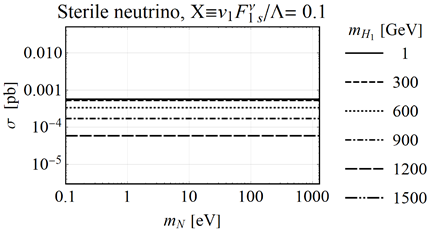}} \
\subfloat[]{\includegraphics[width=7.50cm]{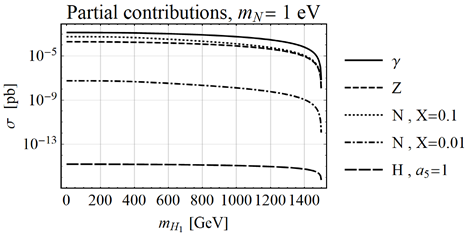}} \\
\subfloat[]{\includegraphics[width=7.25cm]{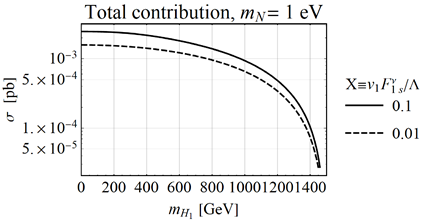}}
\caption{Predictions for the $e^+ e^-\to H^+_1H^-_1$ cross section with $\sqrt{s}=$ 3 TeV. (a): The sterile neutrino contribution for the cross-section as a function of its mass, considering several mass values for the charged scalar. (b): The partial contribution to the cross section as a function of the charged scalar mass. (c): The total cross section as a function of the charged scalar mass.}
	\label{fig:SecoesChoque3TeV}
\end{figure}

In Fig. \ref{fig:SecoesChoque3TeV} are presented the $e^+ e^-\to H^+_1H^-_1$ cross section predictions for $\sqrt{s}=$ 3 TeV. Just like in the previous cases, the top left graph shows that the contribution of the sterile neutrino to the cross section is constant in the range $m_N=[1,1000]$ eV. In the top right graph the partial contributions follow the same hierarchy from Figs. \ref{fig:SecoesChoque380GeV} and \ref{fig:SecoesChoque1500GeV}, where the cross sections are about one order of magnitude smaller than the $\sqrt{s}=$ 1.5 TeV case. At the bottom we show the total cross section, taking into account all diagrams from Fig. \ref{fig:diagramas}. One can notice that for $0\,GeV\,<m_{H_1}<1400\,GeV$ the total cross section stays roughly in the interval $\sigma=$[0.01 pb, $5\times 10^{-5}$ pb], about one order of magnitude below than the $\sqrt{s}=$ 1.5 TeV case.

Overall, we can expect a cross section for the $e^+ e^-\to H^+_1H^-_1$ process roughly in between 0.1 pb and $5 \times 10^{-5}$ pb (depending on $\sqrt{s}$ and on $m_{H_1}$). This is true even if we set $X=a_5=0$, decoupling the sterile neutrino and the Higgs boson from the charged scalars, given that the contributions from the photon and $Z$ boson are predominant in this case. Other Standard Model extensions with charged scalars are likely to give similar results if $H^+ H^- Z$ and $H^+ H^- \gamma$ vertices are present. Also, had we considered other solutions for the scalar mass eigenstates, it would only change the $e^\pm N H^\pm$ and $H H^+ H^-$ vertices, introducing mixing angles that are likely to reduce these vertices. Therefore, changes in these vertices would give negligible differences in our total cross section prediction given that the main contribution comes from the photon, which has its vertex unchanged from choosing other solutions for the scalar mass eigenstates.

Considering the integrated luminosities shown in Table \ref{tab:enegiasCLIC}, for $\sqrt{s}=$380 GeV, where the cross section ranges in between 1 pb and 0.005 pb, the first stage of CLIC should give us a number of events in between $10^6$ and $5 \times 10^3$. Meanwhile, for stage 2, with our cross section predictions ranging from 0.01 pb up to 0.0001 pb, the number of events should be between $2.5 \times 10^4$ and 250. Finally, for $\sqrt{s}=$ 3 TeV, where we predict $\sigma=$[0.01 pb, $5\times 10^{-5}$ pb], the number of events should be between $5\times 10^4$ and 50.

\subsection{The $H_1^\pm$ decay}

At first, one would expect to detect these exotic charged scalars through their decay products, which would leave their signals in the detector. The main decay for $H_1^+$ is $H_1^+ \rightarrow \overline{e} N$, which leads to $e\overline{e}\to e\overline{e}N\overline{N}$ in an electron-positron collider, where the main background for this process in the SM comes from $e\overline{e}\to e\bar{e}\nu_e\overline{\nu}_e$.

At a glance, one would expect the $H_1^+ \rightarrow \overline{e} N H$ and $H_1^+ \rightarrow W^+ H$ decays as well (where $H$ is the SM higgs boson), however, there are no vertices involving $H_1^\pm$ and $H$, which forbids these processes. This can also be seen from Table \ref{table1}, the exotic particles are charged under the $S_3$ and $Z_2$ symmetries, while the SM particles aren't, therefore forbidding these decays.

In the SM $\sigma(e\overline{e}\to e\bar{e}\nu_e\overline{\nu}_e)_\textrm{SM}=4.41\times10^{-1}$, $5.00\times10^{-1}$, $1$ pb for the center of mass energies $\sqrt{s}=380$, 1500, 3000~GeV, while in our $S_3$ model
$\sigma(e\overline{e}\to e\overline{e}N\overline{N})_{S3}\sim10^{-24}$, $\sim10^{-27}$, $\sim10^{-28}$ pb for the respective $\sqrt{s}=380$, 1500, 3000~GeV, considering $m_N=1$~eV, $m_{H_1^+}= 1$~GeV and $X_1=\nu_1 F_{1,s}^\nu/\Lambda=0.1$. Also, the larger $m_{H_1^+}$, the smaller the cross section, making the contribution of the $H_1^\pm$ scalar even smaller for values above 1 GeV. Therefore, the $H_1^\pm$ decay should cause a very small change to the measured $\sigma(e\overline{e}\to e\bar{e}\nu_e\overline{\nu}_e)$ cross section (given that the neutrinos are accounted as missing energy), making them undetectable in this channel \cite{Sicking:2016zcl,Dannheim:2013ypa}.

\begin{center}
	\begin{figure}[!h]
		\subfloat[]{\includegraphics[width=8cm]{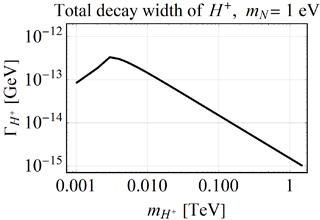}} \qquad
		\subfloat[]{\includegraphics[width=8cm]{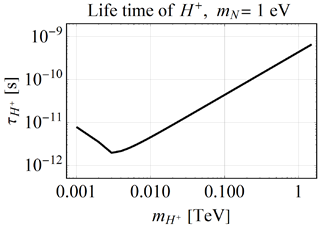}} \\
		\caption{(a) Total decay width of $H_1^+$. (b) Life time of $H_1^+$. In these plots $X_1=\nu_1 F_{1,s}^\nu/\Lambda=0.1$.}
		\label{FIGURE-plots}
	\end{figure}
\end{center}

From the lifetimes shown in Fig. \ref{FIGURE-plots}, we expect the charged scalar to travel in between $2\times 10^{-12}s\cdot c$ up to $7\times 10^{-10}s\cdot c$, equivalent to 0.6 mm up to 21 cm. Therefore, given that the $H_1^+ \rightarrow \overline{e} N$ signal would be overwhelmed by the $e\overline{e}\to e\bar{e}\nu_e\overline{\nu}_e$ process, and the possibility of a measurable track in the detector (since the $H_1^\pm$ is electrically charged), we expect these exotic scalars to be detected directly from their tracks in the detector, instead of being detected through their decay products. The innermost tracker layer is at 165mm from the center \cite{CLICdesignreport}, in this case the travel time must be at least $5.504 \times 10^{-10}$ s, which corresponds to a mass of roughly 1 TeV according to Fig. \ref{FIGURE-plots}.

\section{Conclusions}\label{sec:conclusions}

We have calculated the cross-section at tree-level for the production of charged scalars in eletron-positron collisions in the $S_3 \otimes \mathbb{Z}_2$ model. For our numerical results we considered the Compact Linear Collider energies and luminosities, so we could have a realistic view of possible experimental results. We have also shown the scalar mass eigenstates and interactions, necessary to our calculations.

Out of the exotic particles involved in the diagrams for the cross-section, the sterile neutrino gives a constant contribution to the final result, regardless of its mass. Its coupling constant ($X= F^\nu_{1S} v_1/\Lambda$) to the charged fermions and scalars does have an influence on the final results, but still is a small contribution when compared to the one given by the diagram with the photon. However, a parameter that noticeably changes the cross-section values is the charged scalar mass, leading to a difference of orders of magnitude when comparing the lowest and highest possible cross-section values. In the end, the expected number of events depend heavily on the charged scalar mass, ranging from $10^6$ all the way down to 250. Therefore, this mass value can make the difference in between an observable and a non-observable production of these exotic particles. On top of that, given the small cross section of  $H_1^+ \rightarrow \overline{e^c} N_S$, the fact that its signal would be overwhelmed by the SM background from the  $e\overline{e}\to e\bar{e}\nu_e\overline{\nu}_e$ process, and the possibility of the charged scalar leaving measurable tracks in the detector, we expect this particle to be detected directly in case it exists and is heavy enough (i.e. survives long enough) to reach the innermost tracker.

\section*{Acknowledgements}
G. De Conto would like to thank Coordena\c{c}\~{a}o de Aperfei\c{c}oamento de Pessoal de N\'{i}vel Superior (CAPES) for financial support. J.~M. thanks C\'atedras CONACYT project 1753.

\appendix

\section{Diagonalization solutions for the $3\times3$ CP-even mass matrix}\label{sec:SolucoesCPpar3x3}

In this appendix we show all the solutions for the $3\times3$ CP-even mass matrix discussed in Sec. \ref{sec:CPpar3x3}, except the ones where $v_1=v_2=0$. In the first subsection we consider the vacuum stability condition from Eq. \ref{eq:solderivadas2}, and in the second the condition from Eq. \ref{eq:solderivadas3}. All solutions are presented in the same format: first we show the relations that the parameters must obey, then the diagonalized matrix with the masses squared, and finally the orthogonal diagonalization matrix.

\subsection[Using $v_2= -v_1$]{Using $v_2= -v_1$, $\mu_\zeta^2= \frac{1}{2} \left(-b_1 v_{SM}^2-4 v_1^2 (c_1+c_2)+\mu_{12}^2\right)$, and $\mu_{SM}=-a_4 v_{SM}^2-b_1 v_1^2$.}

\begin{itemize}
	
	\item Solution 1: $\left\{a_4\to \frac{2 v_1^2 (c_1+c_2)}{v_{SM}^2},b_1\to 0,sin \theta \to -\sqrt{1-cos \theta ^2},\mu_{12}^2\to 4 v_1^2 (c_1+c_2)\right\}$\subitem $ diag(m^2_H, m^2_{H_1^0}, m^2_{H_2^0})=\left(
	\begin{array}{ccc}
	4 v_1^2 (c_1+c_2) & 0 & 0 \\
	0 & 4 v_1^2 (c_1+c_2) & 0 \\
	0 & 0 & 4 v_1^2 (c_1+c_2) \\
	\end{array}
	\right)$\subitem $R_{E2}=\left(
	\begin{array}{ccc}
	1 & 0 & 0 \\
	0 & cos \theta  & -\sqrt{1-cos \theta ^2} \\
	0 & \sqrt{1-cos \theta ^2} & cos \theta  \\
	\end{array}
	\right)$
	
	\item Solution 2: $\left\{a_4\to \frac{2 v_1^2 (c_1+c_2)}{v_{SM}^2},b_1\to 0,sin \theta \to \sqrt{1-cos \theta ^2},\mu_{12}^2\to 4 v_1^2 (c_1+c_2)\right\}$\subitem $ diag(m^2_H, m^2_{H_1^0}, m^2_{H_2^0})=\left(
	\begin{array}{ccc}
	4 v_1^2 (c_1+c_2) & 0 & 0 \\
	0 & 4 v_1^2 (c_1+c_2) & 0 \\
	0 & 0 & 4 v_1^2 (c_1+c_2) \\
	\end{array}
	\right)$\subitem $R_{E2}=\left(
	\begin{array}{ccc}
	1 & 0 & 0 \\
	0 & cos \theta  & \sqrt{1-cos \theta ^2} \\
	0 & -\sqrt{1-cos \theta ^2} & cos \theta  \\
	\end{array}
	\right)$
	
	\item Solution 3: $\left\{b_1\to 0,cos \theta \to -\frac{1}{\sqrt{2}},sin \theta \to -\frac{1}{\sqrt{2}},\mu_{12}^2\to 2 a_4 v_{SM}^2\right\}$\subitem $ diag(m^2_H, m^2_{H_1^0}, m^2_{H_2^0})=\left(
	\begin{array}{ccc}
	2 a_4 v_{SM}^2 & 0 & 0 \\
	0 & 4 v_1^2 (c_1+c_2) & 0 \\
	0 & 0 & 2 a_4 v_{SM}^2 \\
	\end{array}
	\right)$\subitem $R_{E2}=\left(
	\begin{array}{ccc}
	1 & 0 & 0 \\
	0 & -\frac{1}{\sqrt{2}} & -\frac{1}{\sqrt{2}} \\
	0 & \frac{1}{\sqrt{2}} & -\frac{1}{\sqrt{2}} \\
	\end{array}
	\right)$
	
	\item Solution 4: $\left\{b_1\to 0,cos \theta \to \frac{1}{\sqrt{2}},sin \theta \to -\frac{1}{\sqrt{2}},\mu_{12}^2\to 2 a_4 v_{SM}^2\right\}$\subitem $ diag(m^2_H, m^2_{H_1^0}, m^2_{H_2^0})=\left(
	\begin{array}{ccc}
	2 a_4 v_{SM}^2 & 0 & 0 \\
	0 & 2 a_4 v_{SM}^2 & 0 \\
	0 & 0 & 4 v_1^2 (c_1+c_2) \\
	\end{array}
	\right)$\subitem $R_{E2}=\left(
	\begin{array}{ccc}
	1 & 0 & 0 \\
	0 & \frac{1}{\sqrt{2}} & -\frac{1}{\sqrt{2}} \\
	0 & \frac{1}{\sqrt{2}} & \frac{1}{\sqrt{2}} \\
	\end{array}
	\right)$
	
	\item Solution 5: $\left\{b_1\to 0,cos \theta \to -\frac{1}{\sqrt{2}},sin \theta \to \frac{1}{\sqrt{2}},\mu_{12}^2\to 2 a_4 v_{SM}^2\right\}$\subitem $ diag(m^2_H, m^2_{H_1^0}, m^2_{H_2^0})=\left(
	\begin{array}{ccc}
	2 a_4 v_{SM}^2 & 0 & 0 \\
	0 & 2 a_4 v_{SM}^2 & 0 \\
	0 & 0 & 4 v_1^2 (c_1+c_2) \\
	\end{array}
	\right)$\subitem $R_{E2}=\left(
	\begin{array}{ccc}
	1 & 0 & 0 \\
	0 & -\frac{1}{\sqrt{2}} & \frac{1}{\sqrt{2}} \\
	0 & -\frac{1}{\sqrt{2}} & -\frac{1}{\sqrt{2}} \\
	\end{array}
	\right)$
	
	\item Solution 6: $\left\{b_1\to 0,cos \theta \to \frac{1}{\sqrt{2}},sin \theta \to \frac{1}{\sqrt{2}},\mu_{12}^2\to 2 a_4 v_{SM}^2\right\}$\subitem $ diag(m^2_H, m^2_{H_1^0}, m^2_{H_2^0})=\left(
	\begin{array}{ccc}
	2 a_4 v_{SM}^2 & 0 & 0 \\
	0 & 4 v_1^2 (c_1+c_2) & 0 \\
	0 & 0 & 2 a_4 v_{SM}^2 \\
	\end{array}
	\right)$\subitem $R_{E2}=\left(
	\begin{array}{ccc}
	1 & 0 & 0 \\
	0 & \frac{1}{\sqrt{2}} & \frac{1}{\sqrt{2}} \\
	0 & -\frac{1}{\sqrt{2}} & \frac{1}{\sqrt{2}} \\
	\end{array}
	\right)$
	
	\item Solution 7: $\left\{a_4\to \frac{2 v_1^2 (c_1+c_2)}{v_{SM}^2},b_1\to 0,cos \theta \to -\frac{1}{\sqrt{2}},sin \theta \to -\frac{1}{\sqrt{2}},\mu_{12}^2\to 4 v_1^2 (c_1+c_2)\right\}$\subitem $ diag(m^2_H, m^2_{H_1^0}, m^2_{H_2^0})=\left(
	\begin{array}{ccc}
	4 v_1^2 (c_1+c_2) & 0 & 0 \\
	0 & 4 v_1^2 (c_1+c_2) & 0 \\
	0 & 0 & 4 v_1^2 (c_1+c_2) \\
	\end{array}
	\right)$\subitem $R_{E2}=\left(
	\begin{array}{ccc}
	1 & 0 & 0 \\
	0 & -\frac{1}{\sqrt{2}} & -\frac{1}{\sqrt{2}} \\
	0 & \frac{1}{\sqrt{2}} & -\frac{1}{\sqrt{2}} \\
	\end{array}
	\right)$
	
	\item Solution 8: $\left\{a_4\to \frac{2 v_1^2 (c_1+c_2)}{v_{SM}^2},b_1\to 0,cos \theta \to \frac{1}{\sqrt{2}},sin \theta \to -\frac{1}{\sqrt{2}},\mu_{12}^2\to 4 v_1^2 (c_1+c_2)\right\}$\subitem $ diag(m^2_H, m^2_{H_1^0}, m^2_{H_2^0})=\left(
	\begin{array}{ccc}
	4 v_1^2 (c_1+c_2) & 0 & 0 \\
	0 & 4 v_1^2 (c_1+c_2) & 0 \\
	0 & 0 & 4 v_1^2 (c_1+c_2) \\
	\end{array}
	\right)$\subitem $R_{E2}=\left(
	\begin{array}{ccc}
	1 & 0 & 0 \\
	0 & \frac{1}{\sqrt{2}} & -\frac{1}{\sqrt{2}} \\
	0 & \frac{1}{\sqrt{2}} & \frac{1}{\sqrt{2}} \\
	\end{array}
	\right)$
	
	\item Solution 9: $\left\{a_4\to \frac{2 v_1^2 (c_1+c_2)}{v_{SM}^2},b_1\to 0,cos \theta \to -\frac{1}{\sqrt{2}},sin \theta \to \frac{1}{\sqrt{2}},\mu_{12}^2\to 4 v_1^2 (c_1+c_2)\right\}$\subitem $ diag(m^2_H, m^2_{H_1^0}, m^2_{H_2^0})=\left(
	\begin{array}{ccc}
	4 v_1^2 (c_1+c_2) & 0 & 0 \\
	0 & 4 v_1^2 (c_1+c_2) & 0 \\
	0 & 0 & 4 v_1^2 (c_1+c_2) \\
	\end{array}
	\right)$\subitem $R_{E2}=\left(
	\begin{array}{ccc}
	1 & 0 & 0 \\
	0 & -\frac{1}{\sqrt{2}} & \frac{1}{\sqrt{2}} \\
	0 & -\frac{1}{\sqrt{2}} & -\frac{1}{\sqrt{2}} \\
	\end{array}
	\right)$
	
	\item Solution 10: $\left\{a_4\to \frac{2 v_1^2 (c_1+c_2)}{v_{SM}^2},b_1\to 0,cos \theta \to \frac{1}{\sqrt{2}},sin \theta \to \frac{1}{\sqrt{2}},\mu_{12}^2\to 4 v_1^2 (c_1+c_2)\right\}$\subitem $ diag(m^2_H, m^2_{H_1^0}, m^2_{H_2^0})=\left(
	\begin{array}{ccc}
	4 v_1^2 (c_1+c_2) & 0 & 0 \\
	0 & 4 v_1^2 (c_1+c_2) & 0 \\
	0 & 0 & 4 v_1^2 (c_1+c_2) \\
	\end{array}
	\right)$\subitem $R_{E2}=\left(
	\begin{array}{ccc}
	1 & 0 & 0 \\
	0 & \frac{1}{\sqrt{2}} & \frac{1}{\sqrt{2}} \\
	0 & -\frac{1}{\sqrt{2}} & \frac{1}{\sqrt{2}} \\
	\end{array}
	\right)$
	
\end{itemize}

\subsection[Using $v_2= v_1$]{Using $v_2= v_1$, $\mu_\zeta^2= \frac{1}{2} \left(-b_1 v_{SM}^2-4 v_1^2 (c_1+c_2)-\mu_{12}^2\right)$ and $\mu_{SM}=-a_4 v_{SM}^2-b_1 v_1^2$.}\label{sec:SolucoesCPpar3x3v1=v2}

\begin{itemize}
	
	\item Solution 1: $\left\{a_4\to \frac{2 v_1^2 (c_1+c_2)}{v_{SM}^2},b_1\to 0,sin \theta \to -\sqrt{1-cos \theta ^2},\mu_{12}^2\to -4 v_1^2 (c_1+c_2)\right\}$\subitem $ diag(m^2_H, m^2_{H_1^0}, m^2_{H_2^0})=\left(
	\begin{array}{ccc}
	4 v_1^2 (c_1+c_2) & 0 & 0 \\
	0 & 4 v_1^2 (c_1+c_2) & 0 \\
	0 & 0 & 4 v_1^2 (c_1+c_2) \\
	\end{array}
	\right)$\subitem $R_{E2}=\left(
	\begin{array}{ccc}
	1 & 0 & 0 \\
	0 & cos \theta  & -\sqrt{1-cos \theta ^2} \\
	0 & \sqrt{1-cos \theta ^2} & cos \theta  \\
	\end{array}
	\right)$
	
	\item Solution 2: $\left\{a_4\to \frac{2 v_1^2 (c_1+c_2)}{v_{SM}^2},b_1\to 0,sin \theta \to \sqrt{1-cos \theta ^2},\mu_{12}^2\to -4 v_1^2 (c_1+c_2)\right\}$\subitem $ diag(m^2_H, m^2_{H_1^0}, m^2_{H_2^0})=\left(
	\begin{array}{ccc}
	4 v_1^2 (c_1+c_2) & 0 & 0 \\
	0 & 4 v_1^2 (c_1+c_2) & 0 \\
	0 & 0 & 4 v_1^2 (c_1+c_2) \\
	\end{array}
	\right)$\subitem $R_{E2}=\left(
	\begin{array}{ccc}
	1 & 0 & 0 \\
	0 & cos \theta  & \sqrt{1-cos \theta ^2} \\
	0 & -\sqrt{1-cos \theta ^2} & cos \theta  \\
	\end{array}
	\right)$
	
	\item Solution 3: $\left\{b_1\to 0,cos \theta \to -\frac{1}{\sqrt{2}},sin \theta \to -\frac{1}{\sqrt{2}},\mu_{12}^2\to -2 a_4 v_{SM}^2\right\}$\subitem $ diag(m^2_H, m^2_{H_1^0}, m^2_{H_2^0})=\left(
	\begin{array}{ccc}
	2 a_4 v_{SM}^2 & 0 & 0 \\
	0 & 2 a_4 v_{SM}^2 & 0 \\
	0 & 0 & 4 v_1^2 (c_1+c_2) \\
	\end{array}
	\right)$\subitem $R_{E2}=\left(
	\begin{array}{ccc}
	1 & 0 & 0 \\
	0 & -\frac{1}{\sqrt{2}} & -\frac{1}{\sqrt{2}} \\
	0 & \frac{1}{\sqrt{2}} & -\frac{1}{\sqrt{2}} \\
	\end{array}
	\right)$
	
	\item Solution 4: $\left\{b_1\to 0,cos \theta \to \frac{1}{\sqrt{2}},sin \theta \to -\frac{1}{\sqrt{2}},\mu_{12}^2\to -2 a_4 v_{SM}^2\right\}$\subitem $ diag(m^2_H, m^2_{H_1^0}, m^2_{H_2^0})=\left(
	\begin{array}{ccc}
	2 a_4 v_{SM}^2 & 0 & 0 \\
	0 & 4 v_1^2 (c_1+c_2) & 0 \\
	0 & 0 & 2 a_4 v_{SM}^2 \\
	\end{array}
	\right)$\subitem $R_{E2}=\left(
	\begin{array}{ccc}
	1 & 0 & 0 \\
	0 & \frac{1}{\sqrt{2}} & -\frac{1}{\sqrt{2}} \\
	0 & \frac{1}{\sqrt{2}} & \frac{1}{\sqrt{2}} \\
	\end{array}
	\right)$
	
	\item Solution 5: $\left\{b_1\to 0,cos \theta \to -\frac{1}{\sqrt{2}},sin \theta \to \frac{1}{\sqrt{2}},\mu_{12}^2\to -2 a_4 v_{SM}^2\right\}$\subitem $ diag(m^2_H, m^2_{H_1^0}, m^2_{H_2^0})=\left(
	\begin{array}{ccc}
	2 a_4 v_{SM}^2 & 0 & 0 \\
	0 & 4 v_1^2 (c_1+c_2) & 0 \\
	0 & 0 & 2 a_4 v_{SM}^2 \\
	\end{array}
	\right)$\subitem $R_{E2}=\left(
	\begin{array}{ccc}
	1 & 0 & 0 \\
	0 & -\frac{1}{\sqrt{2}} & \frac{1}{\sqrt{2}} \\
	0 & -\frac{1}{\sqrt{2}} & -\frac{1}{\sqrt{2}} \\
	\end{array}
	\right)$
	
	\item Solution 6: $\left\{b_1\to 0,cos \theta \to \frac{1}{\sqrt{2}},sin \theta \to \frac{1}{\sqrt{2}},\mu_{12}^2\to -2 a_4 v_{SM}^2\right\}$\subitem $ diag(m^2_H, m^2_{H_1^0}, m^2_{H_2^0})=\left(
	\begin{array}{ccc}
	2 a_4 v_{SM}^2 & 0 & 0 \\
	0 & 2 a_4 v_{SM}^2 & 0 \\
	0 & 0 & 4 v_1^2 (c_1+c_2) \\
	\end{array}
	\right)$\subitem $R_{E2}=\left(
	\begin{array}{ccc}
	1 & 0 & 0 \\
	0 & \frac{1}{\sqrt{2}} & \frac{1}{\sqrt{2}} \\
	0 & -\frac{1}{\sqrt{2}} & \frac{1}{\sqrt{2}} \\
	\end{array}
	\right)$
	
	\item Solution 7: $\left\{a_4\to \frac{2 v_1^2 (c_1+c_2)}{v_{SM}^2},b_1\to 0,cos \theta \to -\frac{1}{\sqrt{2}},sin \theta \to -\frac{1}{\sqrt{2}},\mu_{12}^2\to -4 v_1^2 (c_1+c_2)\right\}$\subitem $ diag(m^2_H, m^2_{H_1^0}, m^2_{H_2^0})=\left(
	\begin{array}{ccc}
	4 v_1^2 (c_1+c_2) & 0 & 0 \\
	0 & 4 v_1^2 (c_1+c_2) & 0 \\
	0 & 0 & 4 v_1^2 (c_1+c_2) \\
	\end{array}
	\right)$\subitem $R_{E2}=\left(
	\begin{array}{ccc}
	1 & 0 & 0 \\
	0 & -\frac{1}{\sqrt{2}} & -\frac{1}{\sqrt{2}} \\
	0 & \frac{1}{\sqrt{2}} & -\frac{1}{\sqrt{2}} \\
	\end{array}
	\right)$
	
	\item Solution 8: $\left\{a_4\to \frac{2 v_1^2 (c_1+c_2)}{v_{SM}^2},b_1\to 0,cos \theta \to \frac{1}{\sqrt{2}},sin \theta \to -\frac{1}{\sqrt{2}},\mu_{12}^2\to -4 v_1^2 (c_1+c_2)\right\}$\subitem $ diag(m^2_H, m^2_{H_1^0}, m^2_{H_2^0})=\left(
	\begin{array}{ccc}
	4 v_1^2 (c_1+c_2) & 0 & 0 \\
	0 & 4 v_1^2 (c_1+c_2) & 0 \\
	0 & 0 & 4 v_1^2 (c_1+c_2) \\
	\end{array}
	\right)$\subitem $R_{E2}=\left(
	\begin{array}{ccc}
	1 & 0 & 0 \\
	0 & \frac{1}{\sqrt{2}} & -\frac{1}{\sqrt{2}} \\
	0 & \frac{1}{\sqrt{2}} & \frac{1}{\sqrt{2}} \\
	\end{array}
	\right)$
	
	\item Solution 9: $\left\{a_4\to \frac{2 v_1^2 (c_1+c_2)}{v_{SM}^2},b_1\to 0,cos \theta \to -\frac{1}{\sqrt{2}},sin \theta \to \frac{1}{\sqrt{2}},\mu_{12}^2\to -4 v_1^2 (c_1+c_2)\right\}$\subitem $ diag(m^2_H, m^2_{H_1^0}, m^2_{H_2^0})=\left(
	\begin{array}{ccc}
	4 v_1^2 (c_1+c_2) & 0 & 0 \\
	0 & 4 v_1^2 (c_1+c_2) & 0 \\
	0 & 0 & 4 v_1^2 (c_1+c_2) \\
	\end{array}
	\right)$\subitem $R_{E2}=\left(
	\begin{array}{ccc}
	1 & 0 & 0 \\
	0 & -\frac{1}{\sqrt{2}} & \frac{1}{\sqrt{2}} \\
	0 & -\frac{1}{\sqrt{2}} & -\frac{1}{\sqrt{2}} \\
	\end{array}
	\right)$
	
	\item Solution 10: $\left\{a_4\to \frac{2 v_1^2 (c_1+c_2)}{v_{SM}^2},b_1\to 0,cos \theta \to \frac{1}{\sqrt{2}},sin \theta \to \frac{1}{\sqrt{2}},\mu_{12}^2\to -4 v_1^2 (c_1+c_2)\right\}$\subitem $ diag(m^2_H, m^2_{H_1^0}, m^2_{H_2^0})=\left(
	\begin{array}{ccc}
	4 v_1^2 (c_1+c_2) & 0 & 0 \\
	0 & 4 v_1^2 (c_1+c_2) & 0 \\
	0 & 0 & 4 v_1^2 (c_1+c_2) \\
	\end{array}
	\right)$\subitem $R_{E2}=\left(
	\begin{array}{ccc}
	1 & 0 & 0 \\
	0 & \frac{1}{\sqrt{2}} & \frac{1}{\sqrt{2}} \\
	0 & -\frac{1}{\sqrt{2}} & \frac{1}{\sqrt{2}} \\
	\end{array}
	\right)$
	
\end{itemize}

\end{document}